# Room-temperature Magnetoelastic Coupling in $UIr_4Al_{15}$

Mingyu Xu[1], Tomasz Klimczuk[2], M. Brian Maple[3], Weiwei Xie[1*]

1. Department of Chemistry, Michigan State University, East Lansing, MI, 48824 USA
2. Faculty of Applied Physics and Mathematics and Advanced Material Center, Gdansk University of Technology, Gdańsk, Poland
3. Department of Physics, University of California, San Diego, CA, 92093 USA

[*]Corresponding author: Dr. Weiwei Xie (xieweiwe@msu.edu)

*Abstract*

**The interplay between electronic and structural degrees of freedom underpins many emergent phenomena in quantum materials, yet magnetoelastic coupling in metallic systems is typically weak and confined to low temperatures or microscopic length scales. Here, we report giant magnetoelastic coupling slightly above room temperature in the uranium-based intermetallic compound $UIr_4Al_{15}$. Using temperature-dependent single-crystal X-ray diffraction together with anisotropic magnetic susceptibility measurements, we directly resolve subtle but reproducible structural distortions coupled to magnetic alignment. Despite the absence of crystallographic symmetry breaking, pronounced anomalies emerge in lattice parameters and selected chemical bond distances near the magnetic transition region, revealing an unusual sensitivity of the crystal structure to magnetic orientation. The coupling enables direct probing of how atomic distances and local chemical bonding govern the electronic and magnetic states in a bulk intermetallic material. Our results establish $UIr_4Al_{15}$ as a rare platform in which magnetism and lattice distortions are strongly intertwined near room temperature, opening new opportunities for magnetically responsive quantum materials and functional magnetic sensing applications.**

Magnetoelastic coupling arises because magnetic interactions depend strongly on atomic distances and bonding geometry[1–6]. When magnetic moments align or reorient, they induce subtle lattice distortions and atomic displacements that modify orbital overlap, electronic hybridization, and magnetic anisotropy. This feedback between magnetism and local structure can produce structural and magnetic responses even in the absence of crystallographic symmetry breaking. Such exchange-driven magnetostriction has been observed in systems ranging from elemental Ni and amorphous metallic alloy [7–10] to two-dimensional van der Waals ferromagnets [11–13] and itinerant ruthenates [14–16] with strong spin-orbit coupling, such as $Sr_4Ru_3O_{10}$ [1,16–18]. In elemental ferromagnets such as Nickel ($T_c$ ~ 627 K) and Iron ($T_c$ ~ 1043 K), magnetoelastic coupling originates mainly from exchange interactions and spin-orbit coupling accompanying ferromagnetic ordering [19,20]. Giant magnetoelastic alloys such as Terfenol-D [21,22] and Galfenol [23,24] exhibit enhanced responses due to strong spin-orbit coupling and field-induced domain reorientation. In itinerant metamagnetic compounds such as FeRh and $Sr_4Ru_3O_{10}$, magnetoelastic effects are closely tied to magnetic phase transitions, meta-magnetism, or magnetic fluctuations that strongly modify lattice parameters and electronic bandwidths [25,26].

Here, we report a new mechanism for magnetoelastic coupling driven by subtle changes in chemical bonding. Unlike conventional magnetostriction, which is typically associated with magnetic ordering or macroscopic strain, this mechanism originates from the sensitivity of magnetic interactions to local interatomic distances and orbital hybridization. By tuning the bonding distances within the crystal lattice, the magnetic susceptibility changes accordingly, revealing a direct coupling between local structure and magnetic response. Such behavior is observed in bulk $UIr_4Al_{15}$ crystals, where temperature-dependent single-crystal X-ray diffraction reveals subtle variations in U-Al bonding distances and anisotropic atomic displacements near 320-330 K, accompanied by strong field-direction-dependent magnetic susceptibility. Remarkably, these changes occur without crystallographic symmetry breaking, indicating an iso-symmetric magnetoelastic transition governed by local bonding rearrangements rather than a conventional magnetic phase transition.

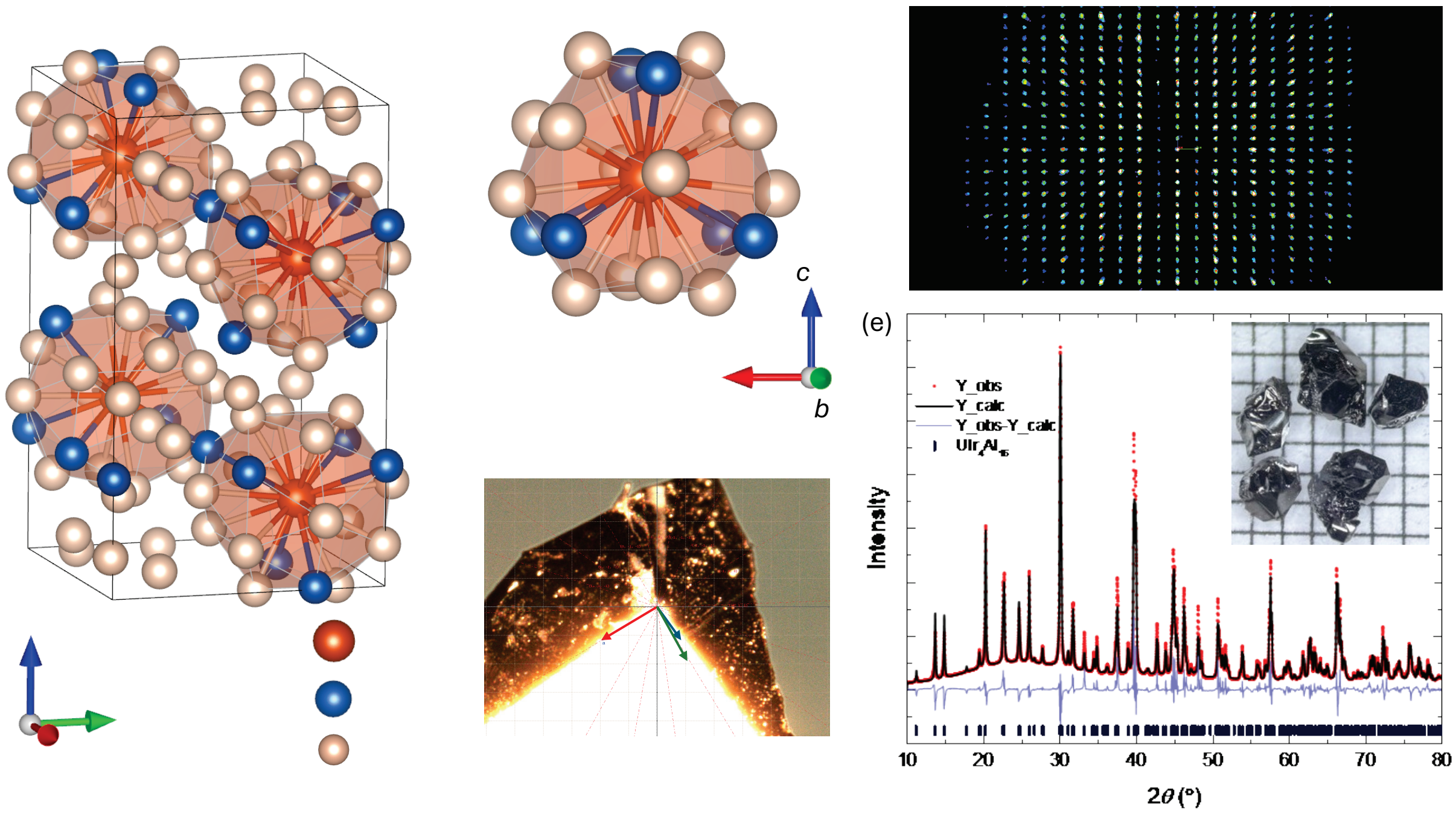

**Figure 1**. Crystal structure and structural characterization of $UIr_4Al_{15}$. (a) Crystal structure of $UIr_4Al_{15}$. (b) Local coordination environment of U within the Ir/Al polyhedral cage. (c) Orientation determination of a single crystal by single-crystal X-ray diffraction. The same crystal was subsequently used for magnetic susceptibility measurements. (d) Reconstructed 0kl reciprocal-space plane obtained from single-crystal X-ray diffraction at 300(2) K. (e) Rietveld refinement of powder X-ray diffraction data confirming phase purity. Insets show optical images of as-grown single crystals on millimeter-scale grid paper.

$UIr_4Al_{15}$ crystallizes in the tetragonal $P4_2/nmc$ space group, consistent with other $RT_4Al_{15}$ compounds ($R$ = U, La, Ce, Pr, Sm, and Gd; $T$ = Rh and Ir) [27,28]. As shown in **Fig. 1*a***, the structure contains distorted U-centered $Ir_6$ polyhedra embedded within an extended Al framework. These polyhedra are connected through corner-sharing Ir atoms, forming quasi-layered structural motifs within the three-dimensional lattice. The local coordination environment of U is illustrated in **Fig. 1*b***. Notably, the four bridging U-Ir bonds associated with corner-sharing connections are significantly longer (3.4776 Å) than the two non-bridging U-Ir bonds (3.2440 Å), producing an almost linear U-Ir-U linkage angle of 179.1°. Such anisotropic bonding geometry suggests highly directional hybridization pathways between U 5*f* and Ir states. The orientation matrix of the single crystal, determined by single-crystal X-ray diffraction and used for anisotropic magnetic susceptibility measurements, is illustrated in **Fig. 1*c***. The reconstructed 0*kl* reciprocal-space plane collected at 300(2) K (**Fig. 1*d***) exhibits sharp and well-defined reflections without diffuse scattering or extra peaks, confirming the absence of twinning or secondary structural modulations. Rietveld refinement of powder X-ray diffraction data (**Fig. 1*e***) confirms the phase purity of the bulk sample, with no detectable impurity phases within the experimental resolution. Energy-dispersive X-ray spectroscopy further exam the stoichiometric composition of $UIr_4Al_{15}$ (Supplementary **Fig. S1**). Its semiquantitative composition should not be interpreted as independent confirmation of exact 1:4:15 stoichiometry.

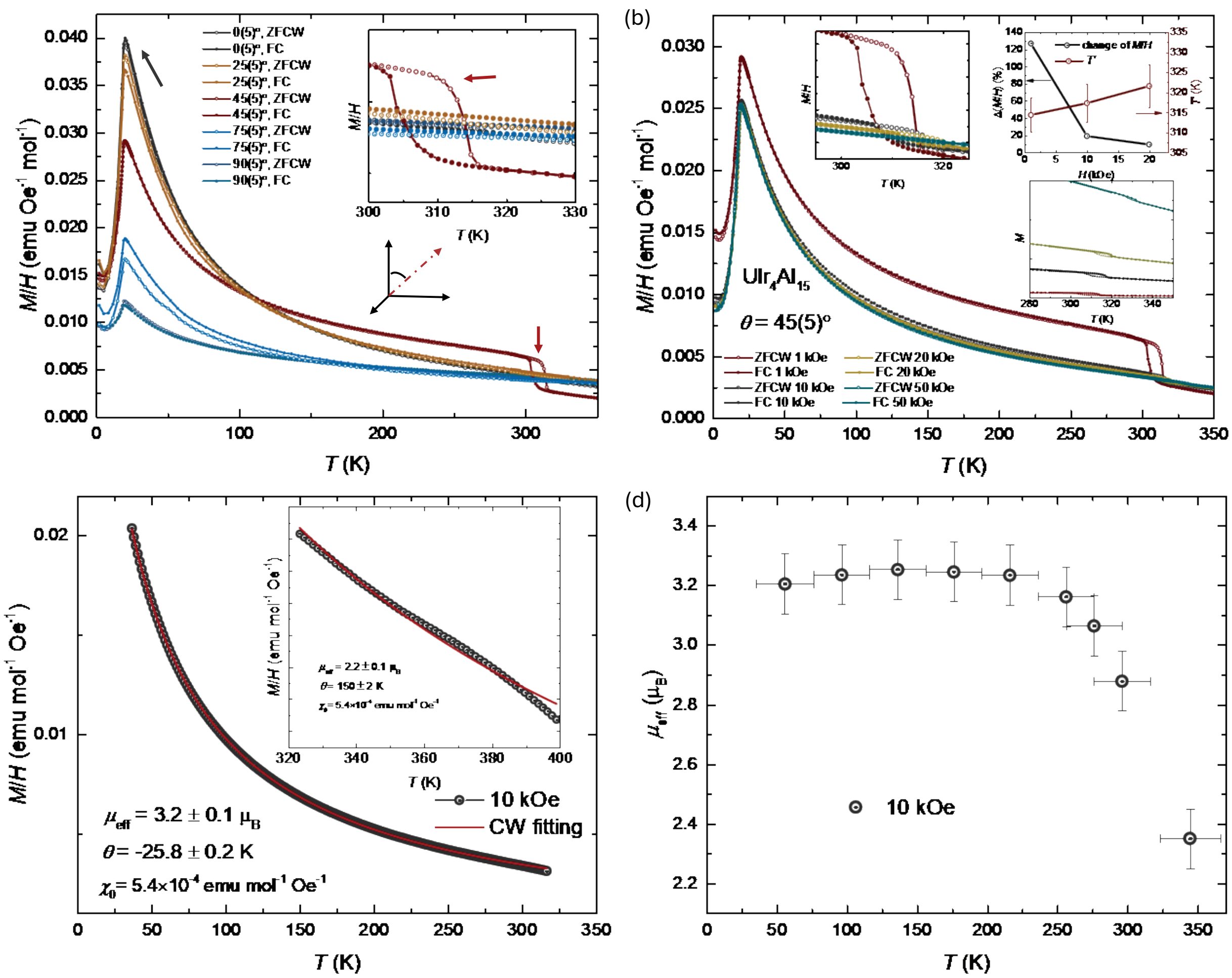

**Figure 2.** Magnetic properties of $UIr_4Al_{15}$. (***a***) Temperature-dependent magnetic susceptibility ($M/H$) measured under zero-field-cooled warming (ZFCW) and field-cooled (FC) conditions for different magnetic-field orientations within the crystallographic *ac*-plane. $T_N$ denotes the antiferromagnetic transition temperature, while $T'$ marks the higher-temperature anomaly near 310 K. The magnetic-field orientation angle $\theta$, defined relative to the crystallographic *c*-axis, is illustrated in the lower inset. The upper-right inset highlights the anisotropic magnetic response between 300 and 320 K. (***b***) Temperature-dependent magnetic susceptibility measured under different magnetic fields, with the field applied at $\theta = 45(5)°$ from the *c*-axis. The left inset enlarges the temperature region between 290 and 330 K surrounding $T'$. The right-upper inset shows the relative change in magnetic susceptibility, $M/H(\Delta(M/H) = (M(T'_{offset})-M(T'_{onset}))/M(T'_{offset})$, together with the evolution of $T'$ as a function of magnetic field. The right-lower inset shows magnetization changes. (***c***, ***d***) and Curie-Weiss fitting of the ZFCW magnetic susceptibility measured under 10 *k*Oe. The inset shows the Curie-Weiss fitting performed in the temperature range above $T'$ and below 400 K. Effective magnetic moments ($\mu_{eff}$) extracted from Curie-Weiss fits over different temperature windows, revealing temperature-dependent evolution of the magnetic response.

**Fig. 2*a*** shows the temperature-dependent magnetic susceptibility (*M*/*H*) of $UIr_4Al_{15}$ measured under zero-field-cooled warming (ZFCW) and field-cooled (FC) conditions with the magnetic field applied along different directions within the *ac*-plane over the temperature range 2-350 K. The angle θ between the magnetic field and the crystallographic *c*-axis is defined in the lower inset. All measurements reveal an antiferromagnetic transition at $T_N \sim 20$ K, followed by suppressed magnetic susceptibility below $T_N$, consistent with dominant antiferromagnetic interactions. A second anomaly, denoted *T'*, emerges near 310 K only for fields applied close to $\theta = 45(5)°$, indicating a highly anisotropic magnetic response. As highlighted in the inset of **Fig. 2*a***, this anomaly exhibits a clear thermal hysteresis between the ZFCW and FC measurements, suggestive of a magnetoelastic or electronically driven transition. At the orientation for $\theta = 45(5)°$, the characteristic anomaly temperature changes with field, and the feature becomes less pronounced in M/H in **Fig. 2*b***. The absolute magnetization step is approximately field-independent over the measured range, as assessed from the available curves; quantitative uncertainties require consistent baseline extraction. A nearly constant $\Delta M$ is distinct from a field-independent susceptibility change, for which $\Delta M = H\Delta\chi$. The high-temperature feature was also reproduced after remounting the same crystal: VSM measurements used a temperature sweep of 2 K $min^{-1}$, whereas the MPMS3 repeat used no-overshoot temperature control with a 2 s wait after the instrument reported temperature stability (**Fig. S2**). Neither data set was vertically shifted or background subtracted. Increasing magnetic field progressively suppresses the magnitude of the anomaly while simultaneously shifting *T'* to higher temperatures. The left inset highlights the evolution of the transition region between 290 and 330 K. The right inset summarizes the field dependence of the relative susceptibility change, defined as $\Delta(M/H) = (M(T'_{offset})-M(T'_{onset}))/\ M(T'_{offset})$. The magnitude of the anomaly decreases dramatically from approximately 130% under 1 kOe to below 10% under 20 kOe, whereas *T'* increases nearly linearly with magnetic field. Such strong field sensitivity suggests an intimate coupling between magnetic anisotropy and the underlying lattice or bonding environment.

To further understand the magnetic behavior, the susceptibility measured at 10 kOe and $\theta = 45(5)°$ was analyzed using a modified Curie-Weiss expression, $\chi = \frac{C}{T-\theta} + \chi_0$, where $\chi \approx M/H$. As shown in **Fig. 2*c***, fitting the ZFCW data over the temperature range 36-316 K yields an effective magnetic moment, $\mu_{eff} = (3.2 \pm 0.1)\ \mu_B$ and a Weiss temperature, -25.8(2) K, indicating dominant antiferromagnetic exchange interactions with an energy scale comparable to $T_N$. The inset of **Fig. 2*c*** shows the Curie-Weiss fitting above *T'*, where the susceptibility deviates more strongly from linear behavior, suggesting modifications to the electronic or magnetic state near the high-temperature anomaly. **Fig. 2*d*** summarizes the evolution of $\mu_{eff}$ extracted from different fitting windows. The effective moment remains nearly constant at ~3.2 $\mu_B$, below approximately 250 K, but decreases substantially when the fitting range includes temperatures above *T'*. To reduce the correlation of CW fitting of $C$, $\theta$, $\chi_0$, the $\chi_0$ is fixed during the individual range of CW fitting, since $\chi_0$ is a temperature-independent variable. This behavior suggests that the anomaly near 310 K is closely linked to changes in magnetic correlations and electronic hybridization, consistent with the proposed bond-driven magnetoelastic coupling in $UIr_4Al_{15}$.

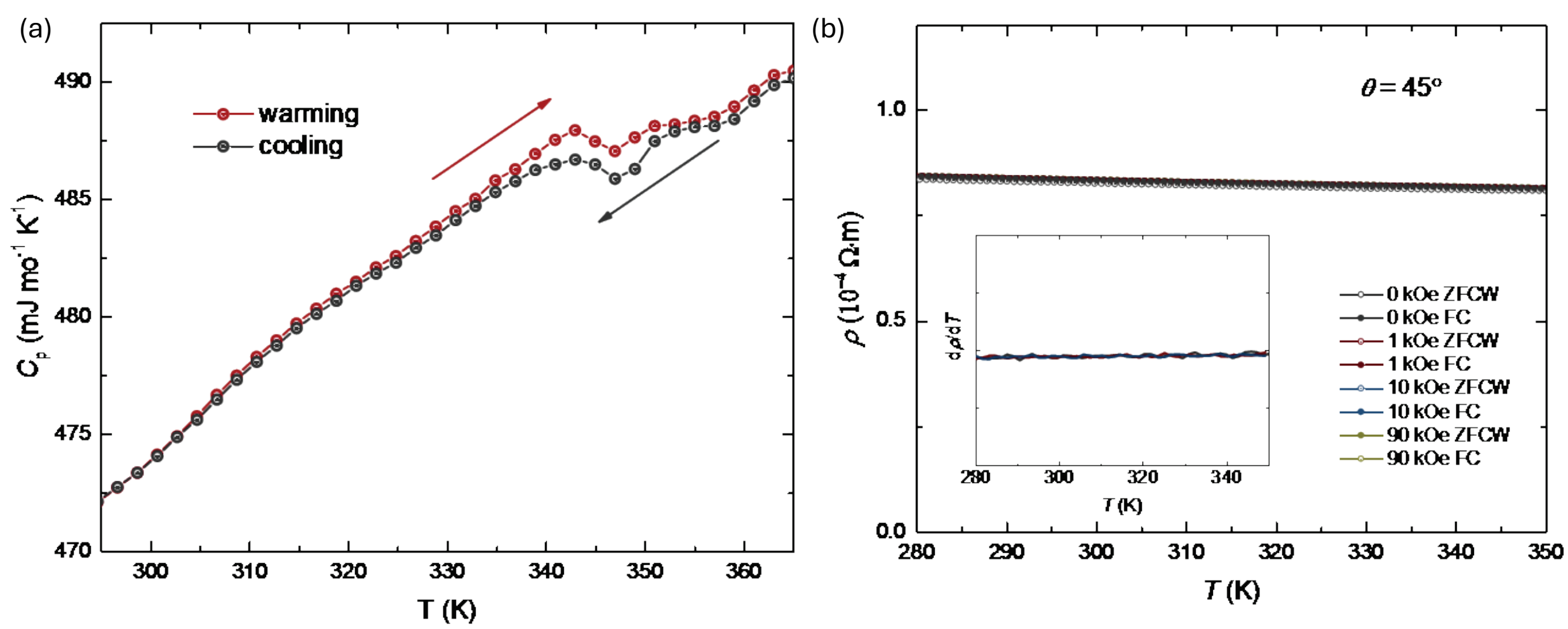

**Figure 3.** Specific heat and resistivity of $UIr_4Al_{15}$ in the temperature range above 300 K. **(*a*)** Temperature-dependent specific heat, $C_p$(T), measured under zero magnetic field. **(*b*)** Temperature-dependent electrical resistivity measured under magnetic fields of 0, 1, 10, and 90 kOe using zero-field-cooled warming (ZFCW) and field-cooled (FC) protocols. The inset shows the temperature derivative of the electrical resistivity, $d\rho/dT$, under different magnetic fields.

**Fig. 3** presents the heat capacity and electrical transport of $UIr_4Al_{15}$ above room temperature. The zero-field heat-capacity data show a broad feature near 310 K; however, there is a clearer transition near 340 K, as shown in **Fig. 3*a***. In contrast, the electrical-transport curves evolve smoothly through the temperature interval of the magnetic anomaly, with no clearly resolved feature under the measured fields (**Fig. 3*b***).

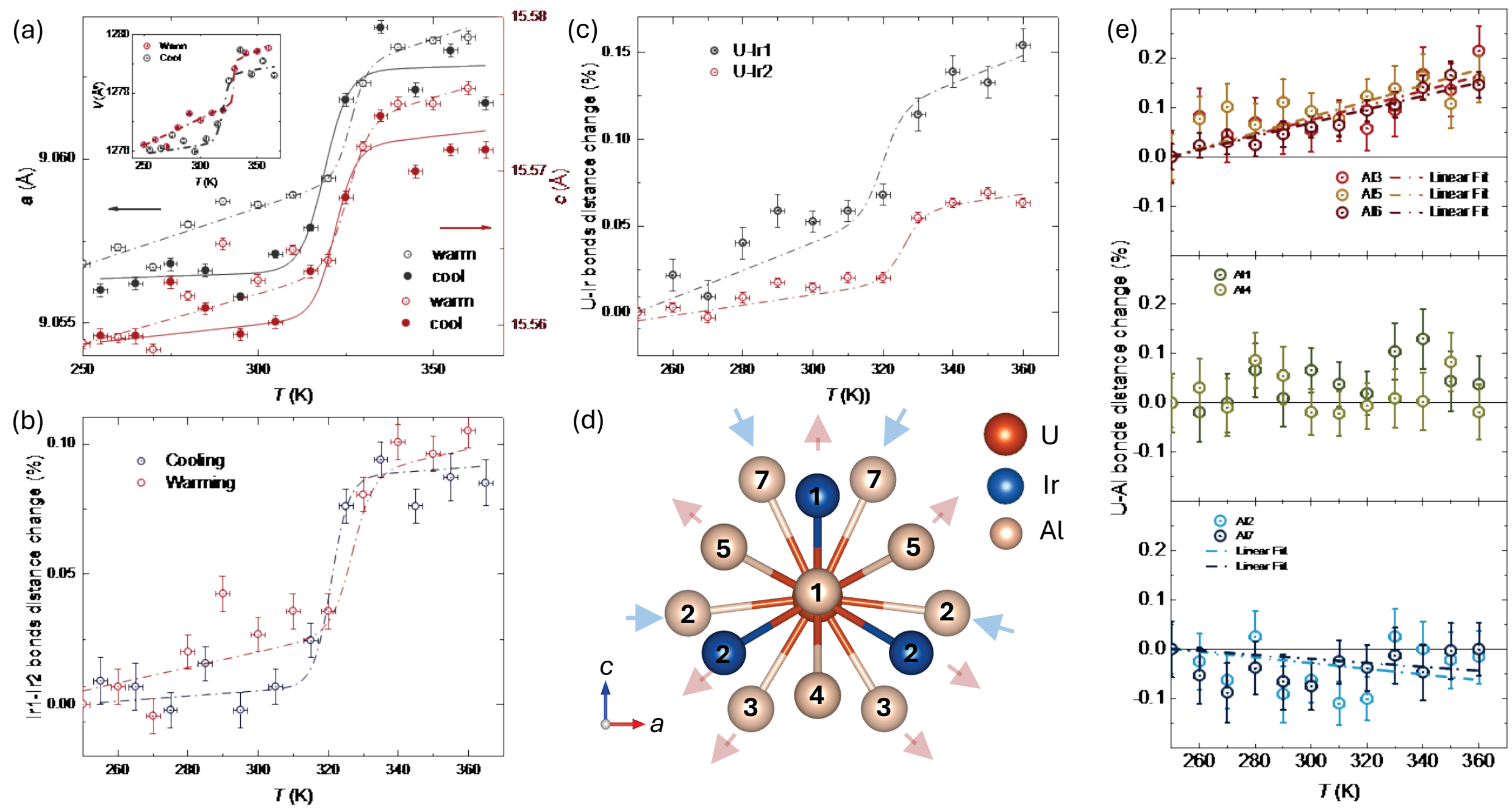

**Figure 4.** Temperature-dependent single-crystal X-ray diffraction analysis of $UIr_4Al_{15}$. **(*a*)** Temperature dependence of the lattice parameters $a$and $c$measured during warming and cooling between 250 and 370 K. Inset shows the corresponding relative change in unit-cell volume, $V/V_0$, over the same temperature range. **(*b*)** Relative change in the Ir1-Ir2 bond distance as a function of temperature during warming and cooling. The abrupt increase near 320-330 K coincides with the magnetic anomaly at $T'$. **(*c*)** Temperature dependence of the relative changes in U-Ir1 and U-Ir2 bond distances. The two inequivalent U-Ir bonds exhibit distinct temperature responses across the transition region, indicating anisotropic local structural distortions. (Solid and dashed-dotted lines in A, B, and C are fitted using $y = C + A(T - 300) + B[1 + \tanh((T - T_0)/\omega)]$.) **(*d*)** Schematic illustration of the temperature-dependent evolution of the U-centered coordination environment, highlighting anisotropic changes in U-Ir and U-Al bond distances across the transition region. Arrows indicate the relative expansion and contraction of selected bonding directions. **(*e*)** Relative changes in inequivalent U-Al bond distances as a function of temperature. Different U-Al bonds exhibit markedly different temperature dependences, revealing anisotropic distortions of the Al framework surrounding U. Dashed lines indicate linear fits to the data.

To further investigate the anomaly near 310-320 K, temperature-dependent single-crystal X-ray diffraction measurements were performed between 250 and 370 K. **Fig. 4*a*** summarizes the evolution of the lattice parameters and unit-cell volume (inset) during both warming and cooling cycles. Both lattice parameters and the unit-cell volume exhibit distinct step-like increases near 320 K, coinciding with the temperature range of the magnetic anomaly observed in the susceptibility measurements. The temperature dependence was fitted using the empirical expression $y = C + A(T - 300) + B[1 + \tanh((T - T_0)/\omega)]$, where $A$, $B$, $C$, $T_0$, and $\omega$ are fitting parameters. The fitted curves reproduce the abrupt lattice response near the transition region, consistent with an iso-symmetric structural anomaly. At 300 K, the Ir1-Ir2 bond distance is 4.4677(1) Å. As shown in **Fig. 4*b***, the Ir1-Ir2 separation also exhibits a pronounced step-like increase near 320-330 K, closely tracking the temperature range of the magnetic anomaly. Because the U atom resides within the Ir/Al coordination cage, we further examined the temperature evolution of the local U-centered bonding environment. **Fig. 4*c*** shows that both U-Ir1 and U-Ir2 bond distances display similar abrupt changes across the transition region, although the magnitudes differ for the two inequivalent bonds, indicating anisotropic local distortions within the U-centered cage. **Fig. 4*d*** schematically illustrates the anisotropic evolution of the U-Ir and U-Al bonding network across the transition. In contrast to the U-Ir bonds, the U-Al bond distances shown in **Fig. 4*e*** evolve more gradually with temperature and do not exhibit a comparably sharp step-like anomaly. Instead, different U-Al bonds display distinct temperature dependences, reflecting subtle anisotropic distortions of the surrounding Al framework. The absence of a sharp discontinuity in the U-Al distances may partially arise from the larger positional uncertainties of the lighter Al atoms in the presence of heavy U and Ir scatterers. Linear fits were therefore applied to emphasize the overall temperature trends of the U-Al bonding environment. Collectively, these results reveal that the anomaly near 320 K is associated with subtle but highly anisotropic rearrangements of the local bonding network, strongly supporting a bond-driven magnetoelastic coupling mechanism in $UIr_4Al_{15}$. The close correspondence between the structural evolution and magnetic anomaly near 320 K strongly suggests an intrinsic magnetoelastic origin of the transition.

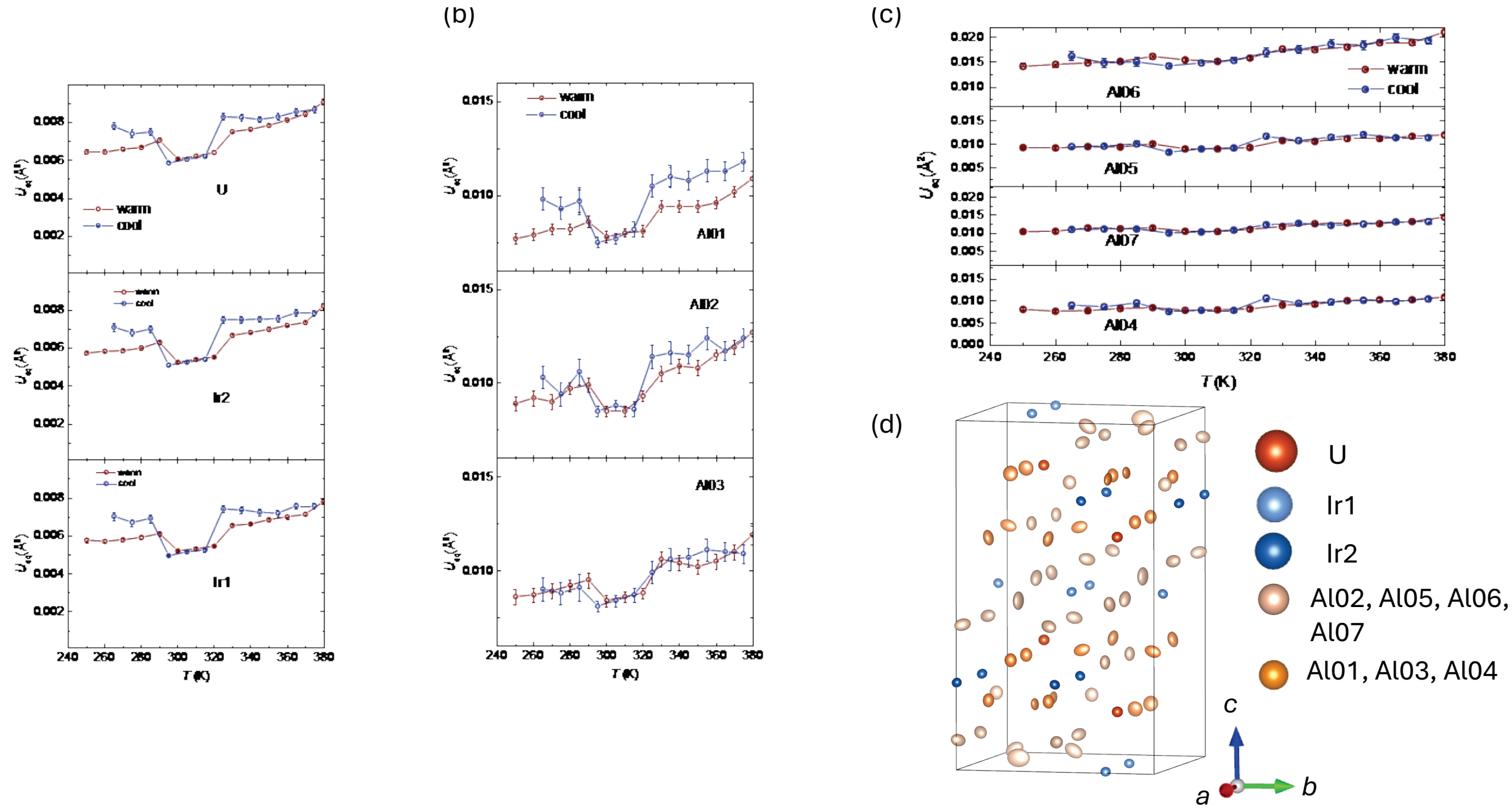

**Figure 5.** Temperature dependence of the equivalent isotropic displacement parameters, ($U_{eq}$), for ***(a)*** U, Ir1, and Ir2; ***(b)*** Al01, Al02, and Al03; and ***(c)*** Al04, Al05, Al06, and Al07. Red and blue symbols represent the warming and cooling cycles, respectively, and the error bars indicate standard uncertainties from the structural refinements. ***(d)*** Unit-cell representation showing the crystallographic U, Ir, and Al sites, with the Al sites color-coded according to the groupings in ***(b)*** and ***(c)***.

The displacement parameters provide a complementary view of the structural evolution in **Fig. 5**. U, both Ir sites, and selected Al (Al01, Al02, and Al03) sites show a nonmonotonic $U_{eq}$ with a low-value interval around 290–320 K on both warming and cooling, followed by recovery on the high-temperature side. Other Al sites (Al04, Al05, Al06, and Al07) show weaker or different temperature dependences. The anisotropic components examined for U and Ir place the principal anomaly in $U_{11}$ and $U_{22}$, without a comparable feature in $U_{33}$ in **Fig. S3**. For the orthogonal tetragonal axes, these components describe displacements along *a*, *b*, and *c*, respectively, and $U_{eq} = (U_{11} + U_{22} + U_{33})/3$. ADPs describe the distribution of atomic positions and can include thermal motion, static disorder, and refinement effects; their decrease suggests phonon hardening or orbital ordering. Taking together, the bond-distance and displacement trends motivate a possible change in the local electronic environment. Modified hybridization involving U 5*f*, Ir 5*d*, and Al states could influence magnetic anisotropy.

In summary, we uncover a reproducible magnetic anomaly at a tilted field orientation near room temperature, together with nonmonotonic displacement parameters and step-like bonding changes in $UIr_4Al_{15}$ near room temperature, where subtle rearrangements of the local bonding environment produce a strong and highly anisotropic magnetic response without crystallographic symmetry breaking. Temperature-dependent single-crystal X-ray diffraction reveals abrupt changes in lattice parameters and selected U-Ir bonding distances near 320-330 K, coinciding with a field-direction-dependent magnetic anomaly. Unlike conventional magnetoelastic systems dominated by long-range magnetic ordering or macroscopic strain, the response in $UIr_4Al_{15}$ originates from atomic-scale distortions within the U-centered Ir/Al coordination cage, highlighting the sensitivity of U-5*f* electronic states to local chemical bonding and hybridization. The strong interplay among bonding, lattice distortions, and magnetic anisotropy establishes a distinct mechanism for magnetoelastic coupling in correlated intermetallic materials. More broadly, our results demonstrate that local bonding engineering can provide a powerful route for controlling magnetic and electronic behavior in bulk quantum materials, opening opportunities for magnetically responsive functional materials and room-temperature magnetic sensing technologies.

## Supporting Information

Experimental details; single crystal X-ray diffraction study; structure and refinement data, SEM-EDS, and Specific heat measurements.

## Author Information

Corresponding Author: xieweiwe@msu.edu

## Acknowledgments

M.X. at Michigan State University, which was supported by the U.S. Department of Energy (DOE), Division of Basic Energy Sciences under Contract DE-SC0023648. W.X. was supported by NSF-DMR-2422361. M.B.M. was supported by the National Nuclear Security Administration (NNSA) under the Stewardship Science Academic Alliance Program through the US DOE under Grant DE-NA0004235.

## References

(1) Marques, C. A.; Rhodes, L. C.; Osmolska, W.; Lane, H.; Benedičič, I.; Naritsuka, M.; Berge, S. A.; Fittipaldi, R.; Lettieri, M.; Vecchione, A.; Wahl, P. Emergent Exchange-Driven Giant Magnetoelastic Coupling in a Correlated Itinerant Ferromagnet. *Nat. Phys.* **2025**, *21* (8), 1243–1249. https://doi.org/10.1038/s41567-025-02893-x.

(2) Jiang, S.; Xie, H.; Shan, J.; Mak, K. F. Exchange Magnetostriction in Two-Dimensional Antiferromagnets. *Nat. Mater.* **2020**, *19* (12), 1295–1299. https://doi.org/10.1038/s41563-020-0712-x.

(3) Canfield, P. C.; Bud'ko, S. L.; Ni, N.; Kreyssig, A.; Goldman, A. I.; McQueeney, R. J.; Torikachvili, M. S.; Argyriou, D. N.; Luke, G.; Yu, W. Structural, Magnetic and Superconducting Phase Transitions in $CaFe_2As_2$ under Ambient and Applied Pressure. *Physica C Supercond.* **2009**, *469* (9–12), 404–412. https://doi.org/10.1016/j.physc.2009.03.033.

(4) Jia, S.; Jiramongkolchai, P.; Suchomel, M. R.; Toby, B. H.; Checkelsky, J. G.; Ong, N. P.; Cava, R. J. Ferromagnetic Quantum Critical Point Induced by Dimer-Breaking in $SrCo_2(Ge_{1-x}P_x)_2$. *Nat. Phys.* **2011**, *7* (3), 207–210. https://doi.org/10.1038/nphys1868.

(5) Callen, E. R.; Callen, H. B. Static Magnetoelastic Coupling in Cubic Crystals. *Phys. Rev.* **1963**, *129* (2), 578–593. https://doi.org/10.1103/PhysRev.129.578.

(6) Hoffmann, R.; Zheng, C. Making and Breaking Bonds in the Solid State: The $ThCr_2Si_2$ Structure. *J. Phys. Chem.* **1985**, *89* (20), 4175–4181. https://doi.org/10.1021/j100266a007.

(7) Azumi, K.; Goldman, J. E. Volume Magnetostriction in Nickel and the Bethe-Slater Interaction Curve. *Phys. Rev.* **1954**, *93* (3), 630–631. https://doi.org/10.1103/PhysRev.93.630.

(8) Gallagher, K. A.; Willard, M. A.; Zabenkin, V. N.; Laughlin, D. E.; McHenry, M. E. Distributed Exchange Interactions and Temperature Dependent Magnetization in Amorphous $Fe_{88-x}Co_xZr_7B_4Cu$ Alloys. *J. Appl. Phys.* **1999**, *85* (8), 5130–5132. https://doi.org/10.1063/1.369100.

(9) Gibbs, M. R. J. Anisotropy and Magnetostriction in Amorphous Alloys. *J. Magn. Magn. Mater.* **1990**, *83* (1–3), 329–333. https://doi.org/10.1016/0304-8853(90)90536-Y.

(10) Spano, M. L.; Hathaway, K. B.; Savage, H. T. Magnetostriction and Magnetic Anisotropy of Field Annealed Metglas* 2605 Alloys via Dc M-H Loop Measurements under Stress. *J. Appl. Phys.* **1982**, *53* (3), 2667–2669. https://doi.org/10.1063/1.330932.

(11) Wang, C.; Zhou, X.; Zhou, L.; Pan, Y.; Lu, Z.-Y.; Wan, X.; Wang, X.; Ji, W. Bethe-Slater-Curve-like Behavior and Interlayer Spin-Exchange Coupling Mechanisms in Two-Dimensional Magnetic Bilayers. *Phys. Rev. B* **2020**, *102* (2), 020402. https://doi.org/10.1103/PhysRevB.102.020402.

(12) Chen, W.; Sun, Z.; Wang, Z.; Gu, L.; Xu, X.; Wu, S.; Gao, C. Direct Observation of van Der Waals Stacking–Dependent Interlayer Magnetism. *Science (1979).* **2019**, *366* (6468), 983–987. https://doi.org/10.1126/science.aav1937.

(13) Vijay, K.; Vavilapalli, D. S.; Arya, A.; Srivastava, S. K.; Singh, R.; Sagdeo, A.; Jha, S. N.; Kumar, K.; Banik, S. Magneto-Strain Effects in 2D Ferromagnetic van Der Waal Material $CrGeTe_3$. *Sci. Rep.* **2023**, *13* (1), 8579. https://doi.org/10.1038/s41598-023-35038-2.

(14) Ohmichi, E.; Yoshida, Y.; Ikeda, S. I.; Shirakawa, N.; Osada, T. Colossal Magnetoresistance Accompanying a Structural Transition in a Highly Two-Dimensional Metallic State of $Ca_3Ru_2O_7$. *Phys. Rev. B* **2004**, *70* (10), 104414. https://doi.org/10.1103/PhysRevB.70.104414.

(15) Naritsuka, M.; Benedičič, I.; Rhodes, L. C.; Marques, C. A.; Trainer, C.; Li, Z.; Komarek, A. C.; Wahl, P. Compass-like Manipulation of Electronic Nematicity in $Sr_3Ru_2O_7$. *Proc. Natl. Acad. Sci. U. S. A.* **2023**, *120* (36). https://doi.org/10.1073/pnas.2308972120.
(16) Gupta, R.; Kim, M.; Barath, H.; Cooper, S. L.; Cao, G. Field- and Pressure-Induced Phases in $Sr_4Ru_3O_{10}$: A Spectroscopic Investigation. *Phys. Rev. Lett.* **2006**, *96* (6), 067004. https://doi.org/10.1103/PhysRevLett.96.067004.
(17) Weickert, F.; Civale, L.; Maiorov, B.; Jaime, M.; Salamon, M. B.; Carleschi, E.; Strydom, A. M.; Fittipaldi, R.; Granata, V.; Vecchione, A. Missing Magnetism in $Sr_4Ru_3O_{10}$: Indication for Antisymmetric Exchange Interaction. *Sci. Rep.* **2017**, *7* (1), 3867. https://doi.org/10.1038/s41598-017-03648-2.
(18) Xu, Z.; Xu, X.; Freitas, R. S.; Long, Z.; Zhou, M.; Fobes, D.; Fang, M.; Schiffer, P.; Mao, Z.; Liu, Y. Magnetic, Electrical Transport, and Thermoelectric Properties of $Sr_4Ru_3O_{10}$: Evidence for a Field-Induced Electronic Phase Transition at Low Temperatures. *Phys. Rev. B* **2007**, *76* (9), 094405. https://doi.org/10.1103/PhysRevB.76.094405.
(19) Jiles, D. C. Theory of the Magnetomechanical Effect. *J. Phys. D Appl. Phys.* **1995**, *28* (8), 1537–1546. https://doi.org/10.1088/0022-3727/28/8/001.
(20) Daalderop, G. H. O.; Kelly, P. J.; Schuurmans, M. F. H. First-Principles Calculation of the Magnetocrystalline Anisotropy Energy of Iron, Cobalt, and Nickel. *Phys. Rev. B* **1990**, *41* (17), 11919–11937. https://doi.org/10.1103/PhysRevB.41.11919.
(21) Clark, A.; Verhoven, J.; McMasters, O.; Gibson, E. Magnetostriction in Twinned [112] Crystals of $Tb_{0.27}Dy_{0.73}Fe_2$. *IEEE Trans. Magn.* **1986**, *22* (5), 973–975. https://doi.org/10.1109/TMAG.1986.1064403.
(22) Jiles, D. C.; Hariharan, S. Interpretation of the Magnetization Mechanism in Terfenol-D Using Barkhausen Pulse-Height Analysis and Irreversible Magnetostriction. *J. Appl. Phys.* **1990**, *67* (9), 5013–5015. https://doi.org/10.1063/1.344707.
(23) Clark, A. E.; Restorff, J. B.; Wun-Fogle, M.; Lograsso, T. A.; Schlagel, D. L. Magnetostrictive Properties of Body-Centered Cubic Fe-Ga and Fe-Ga-Al Alloys. *IEEE Trans. Magn.* **2000**, *36* (5), 3238–3240. https://doi.org/10.1109/20.908752.
(24) Meisenheimer, P. B.; Steinhardt, R. A.; Sung, S. H.; Williams, L. D.; Zhuang, S.; Nowakowski, M. E.; Novakov, S.; Torunbalci, M. M.; Prasad, B.; Zollner, C. J.; Wang, Z.; Dawley, N. M.; Schubert, J.; Hunter, A. H.; Manipatruni, S.; Nikonov, D. E.; Young, I. A.; Chen, L. Q.; Bokor, J.; Bhave, S. A.; Ramesh, R.; Hu, J.-M.; Kioupakis, E.; Hovden, R.; Schlom, D. G.; Heron, J. T. Engineering New Limits to Magnetostriction through Metastability in Iron-Gallium Alloys. *Nat. Commun.* **2021**, *12* (1), 2757. https://doi.org/10.1038/s41467-021-22793-x.
(25) Wolloch, M.; Gruner, M. E.; Keune, W.; Mohn, P.; Redinger, J.; Hofer, F.; Suess, D.; Podloucky, R.; Landers, J.; Salamon, S.; Scheibel, F.; Spoddig, D.; Witte, R.; Roldan Cuenya, B.; Gutfleisch, O.; Hu, M. Y.; Zhao, J.; Toellner, T.; Alp, E. E.; Siewert, M.; Entel, P.; Pentcheva, R.; Wende, H. Impact of Lattice Dynamics on the Phase Stability of Metamagnetic FeRh: Bulk and Thin Films. *Phys. Rev. B* **2016**, *94* (17), 174435. https://doi.org/10.1103/PhysRevB.94.174435.
(26) Bordel, C.; Juraszek, J.; Cooke, D. W.; Baldasseroni, C.; Mankovsky, S.; Minár, J.; Ebert, H.; Moyerman, S.; Fullerton, E. E.; Hellman, F. Fe Spin Reorientation across the Metamagnetic Transition in Strained FeRh Thin Films. *Phys. Rev. Lett.* **2012**, *109* (11), 117201. https://doi.org/10.1103/PhysRevLett.109.117201.

(27) Tobash, P. H.; Scott, B. L.; Sidorov, V. A.; Ronning, F.; Gofryk, K.; Thompson, J. D.; Albers, R. C.; Zhu, J.-X.; Jones, M. D.; Bauer, E. D. Heavy Fermion Behavior in the New Antiferromagnetic Compound $UIr_4Al_{15}$. *J. Phys. Conf. Ser.* **2011**, *273*, 012061. https://doi.org/10.1088/1742-6596/273/1/012061.

(28) Nesterenko, S.; Avzuragova, V.; Tursina, A.; Strydom, A. M. New Ternary Aluminides $RERh_4Al_{15}$ (RE = La, Ce, Pr, Sm, Gd). *J. Alloys Compd.* **2019**, *792*, 1061–1067. https://doi.org/10.1016/j.jallcom.2019.04.026.

# Supplementary Information

## Room-temperature Magnetoelastic Coupling in $UIr_4Al_{15}$

Mingyu Xu[1], Tomasz Klimczuk[2], M. Brian Maple[3], Weiwei Xie[1*]

1. Department of Chemistry, Michigan State University, East Lansing, MI, 48824 USA
2. Faculty of Applied Physics and Mathematics and Advanced Material Center, Gdansk University of Technology, Gdańsk, Poland
3. Department of Physics, University of California, San Diego, CA, 92093 USA

[*]Corresponding author: Dr. Weiwei Xie (xieweiwe@msu.edu)

### Table of Contents

**Experimental Details**

**Crystal Synthesis:** Single crystals of $UIr_4Al_{15}$ were synthesized utilizing a high-temperature solution growth method [1]. Uranium pieces (New Brunswick Laboratory), Iridium powder (Mennica Metale, 99.9%), and Aluminum shot (Alfa Aesar, 99.9% metal basis) with an atomic ratio of 1:1:50 were placed into a Canfield Crucible Set (CCS)[2,3] and subjected to a furnace where the temperature was heated up to 1180 °C. Then, the system underwent a controlled cooling process to 720 °C over 86 hours. At 720 °C, single crystals of $UIr_4Al_{15}$ were separated from the melt using a laboratory centrifuge.

**Ambient Pressure Single-Crystal X-ray diffraction (SCXRD):** A $UIr_4Al_{15}$ single crystal (0.11 × 0.08 × 0.04 $mm^3$) was mounted on a nylon loop with Paratone oil and measured at room temperature, then the same crystal was mounted on a nylon loop with nail polish and measured from 250 K to 370 K using a Rigaku XtaLAB Synergy Dualflex diffractometer with a HyPix detector. Data were collected using ω scans with Mo Kα radiation (λ = 0.71073 Å) from a micro-focus sealed tube (50 kV, 1 mA). Data collection strategies were optimized using CrysAlisPro (v1.171.43.144a). Data reduction included Lorentz and polarization corrections, numerical absorption correction based on Gaussian integration over a multifaceted crystal model[4], and empirical spherical harmonics correction using the SCALE3 ABSPACK algorithm[5]. Structures were solved and refined using Olex2 with the SHELXTL package[6–8]. Refinement details and atomic coordination parameters at 300 K and 350 K are summarized in **Tables 1** and **2**, respectively**.**

**Scanning Electron Microscope and Energy-Dispersive Spectroscopy:** The phase composition was analyzed employing a JEOL 6610LV scanning electron microscope equipped with a tungsten hairpin emitter (JEOL Ltd., Tokyo, Japan). For elemental analysis, energy-dispersive X-ray spectroscopy (EDX) was conducted utilizing an Oxford Instruments AZtec system (Oxford Instruments, High Wycomb, Buckinghamshire, England), operating software version 3.1. This setup included a 20 $mm^2$ Silicon Drift Detector (SDD) and an ultra-thin window integrated with the JEOL 6610LV SEM. The single crystals of $UIr_4Al_{15}$ were attached to carbon adhesive tape and introduced into the SEM chamber for examination at an accelerating voltage of 20 kV. Data acquisition entailed collecting spectra at multiple points along the individual crystals over an optimized timeframe. Quantitative compositional analysis was performed using SEM Quant software, which applies matrix-effect corrections to the intensity measurements.

**Magnetization, Electronic Resistivity, and Specific Heat Measurements:** Temperature- and magnetic-field-dependent magnetization and resistance measurements, as well as temperature-dependent specific heat measurements, were carried out using Quantum Design (QD), Magnetic Property Measurement Systems (MPMS3, max field 50 kOe), and Physical Property Measurement System (DynaCool, max field 90 kOe). Temperature- and field-dependent DC scan and VSM magnetization measurements were taken for $H$ parallel to the crystallographic $a$-axis by placing the SCXRD-characterized sample between two collapsed plastic straws, with the third, uncollapsed, straw providing support as an outer sheath, or by using a quartz sample holder. Other magnetic field direction measurements were made on the PTFE tape on the section of the selected angle-cut straw. Another straw provides support as a sheath on the outside. Samples were fixed on the straw, quartz sample holder, or PTEF tape using GE-7031 varnish. In the VSM magnetization assessments, an oscillation peak amplitude of 4 mm and a mean acquisition time of 2 seconds were employed to ensure precise data collection. DC electrical resistance measurements were performed in a standard four-contact geometry using the PPMS ACT option. 50 μm-diameter Pt wires were bonded to the samples with silver paint (DuPont 4929N), with contact resistances of about 2-3 Ohms. Specific heat capacity measurements under varying temperatures were performed using the relaxation method. N-grease and H-grease were used during measurements across different temperature ranges.

**Single Crystal X-ray Diffraction Study**

To investigate the temperature-structure evolution, temperature-dependent SCXRD measurements were performed from 250 K to 370 K. Selected (300 K and 350 K) refinement parameters of representative temperatures are summarized in **Table 1**. Atomic coordinates refined are listed in **Table 2**. The structure was solved and refined using the Bruker SHELXTL Software Package[6,7], with space group $P4_2/nmc$ and Z = 4 for the formula unit $UIr_4Al_{15}$. No vacancy was observed in the limit resolution of the experiment. The final anisotropic full-matrix least-squares refinement on $F^2$ with 60 variables converged at $R_1$ = 2.51 % at 300 K and $R_1$ = 2.53 % at 350 K, according to the observed data, and $wR_{2\ 300K}$ = 4.80 % and $wR_{2\ 350K}$ = 4.99 % for all data. The goodness-of-fit was 1.055 and 1.079. The largest peak in the final difference electron density synthesis was 2.06 and 2.01 $e^-/Å^3$, and the largest hole was -1.65 and -1.72 $e^-/Å^3$. Based on the final model, the calculated density was 7.343 $g/cm^3$, 2380 $e^-$ of F (000), at 300 K, and 7.328 $g/cm^3$, 2380 $e^-$ of F (000), at 350 K. The crystal structure of $UIr_4Al_{15}$ at different temperatures (from 250K to 370 K) is the same. All of them belong to the $P4_2/nmc$ space group. No symmetry change observed in this temperature range.

**Table S1.** Crystal structure and refinement data for $UIr_4Al_{15}$ at 300 K (ambient pressure) and 350 K. Values in parentheses represent estimated standard deviations from the refinements.

| **Pressure (GPa)** | Ambient | Ambient |
|---|---|---|
| **Temperature (K)** | 300(2) | 350(2) |
| **Space Group** | $P4_2/nmc$ | $P4_2/nmc$ |
| **Unit cell dimensions** | *a* = 9.0603(3) Å, *c* = 15.5532(7) Å. | *a* = 9.06360(10) Å, *c* = 15.5744(4) Å. |
| **Volume (Å$^3$)** | 1276.75(8) | 1279.42(4) |
| **Z** | 4 | 4 |
| **Density (Calculated) (g/m$^3$)** | 7.343 | 7.328 |
| **Absorption coefficient (mm$^{-1}$)** | 55.182 | 55.067 |
| **F(000)** | 2380.0 | 2380.0 |
| **Reflections collected** | 5280 | 18677 |
| **Independent reflections** | 936 [Rint = 0.0480] | 1064 [Rint = 0.0787] |
| **Data/Restraints/Parameters** | 936/0/60 | 1064/0/60 |
| **Final R indices** | $R_1(I \geq 2\sigma(I))$ = 0.0218 $wR_2(I \geq 2\sigma(I))$ = 0.0469; $R_1$(all) = 0.0251, $wR_2$(all) = 0.0480. | $R_1(I \geq 2\sigma(I))$ = 0.0226, $wR_2(I \geq 2\sigma(I))$ = 0.0490; $R_1$(all) = 0.0253, $wR_2$(all) = 0.0499. |
| **Largest diff. peak and hole (e Å$^{-3}$)** | 2.06/-1.65 | 2.01/-1.72 |
| **Goodness-of-fit on F$^2$** | 1.055 | 1.079 |

**Table S2.** Atomic coordinates for $UIr_4Al_{15}$ at 300 K and 350 K.

| **Temperature** | **Atom** | **Wyck.** | ***x*** | ***y*** | ***z*** | **Occ.** | **$U_{eq}$** |
|---|---|---|---|---|---|---|---|
| 300 K | U | 4*d* | ¾ | ¾ | 0.66298(2) | 1 | 0.0054(1) |
| | Ir1 | 8*g* | ¾ | 0.51932(3) | 0.50346(2) | 1 | 0.0044(1) |
| | Ir2 | 8*f* | 0.50209(2) | 0.49791(2) | ¼ | 1 | 0.0044(1) |
| | Al1 | 8*g* | ¾ | 0.40120(20) | 0.6534(1) | 1 | 0.0085(4) |
| | Al2 | 8*g* | ¼ | 0.59840(20) | 0.8122(1) | 1 | 0.0069(4) |
| | Al3 | 8*g* | ¾ | 0.57060(30) | 0.3378(1) | 1 | 0.0071(4) |
| | Al4 | 8*g* | 0.4910(2) | ¼ | 0.6832(1) | 1 | 0.0071(5) |
| | Al5 | 16*h* | 0.4983(2) | 0.96800(20) | 0.5839(1) | 1 | 0.0086(4) |
| | Al6 | 4*c* | ¾ | ¼ | 0.4748(2) | 1 | 0.0141(7) |
| | Al7 | 8*g* | 0.8994(3) | ¾ | 0.4780(1) | 1 | 0.0100(5) |
| 350 K | U | 4*d* | ¾ | ¾ | 0.66300(2) | 1 | 0.0079(1) |
| | Ir1 | 8*g* | ¾ | 0.51930(3) | 0.50346(2) | 1 | 0.0068(1) |
| | Ir2 | 8*f* | 0.50208(2) | 0.49792(2) | ¼ | 1 | 0.0070(1) |
| | Al1 | 8*g* | ¾ | 0.4018(2) | 0.6534(1) | 1 | 0.0108(4) |
| | Al2 | 8*g* | ¼ | 0.5979(2) | 0.8121(1) | 1 | 0.0094(3) |
| | Al3 | 8*g* | ¾ | 0.5705(2) | 0.3377(1) | 1 | 0.0102(4) |
| | Al4 | 8*g* | 0.4914(2) | ¼ | 0.6829(1) | 1 | 0.0100(4) |
| | Al5 | 16*h* | 0.4986(1) | 0.9680(2) | 0.5839(1) | 1 | 0.0112(3) |
| | Al6 | 4*c* | ¾ | ¼ | 0.4754(2) | 1 | 0.0181(6) |
| | Al7 | 8*g* | 0.8997(2) | ¾ | 0.4780(1) | 1 | 0.0128(4) |

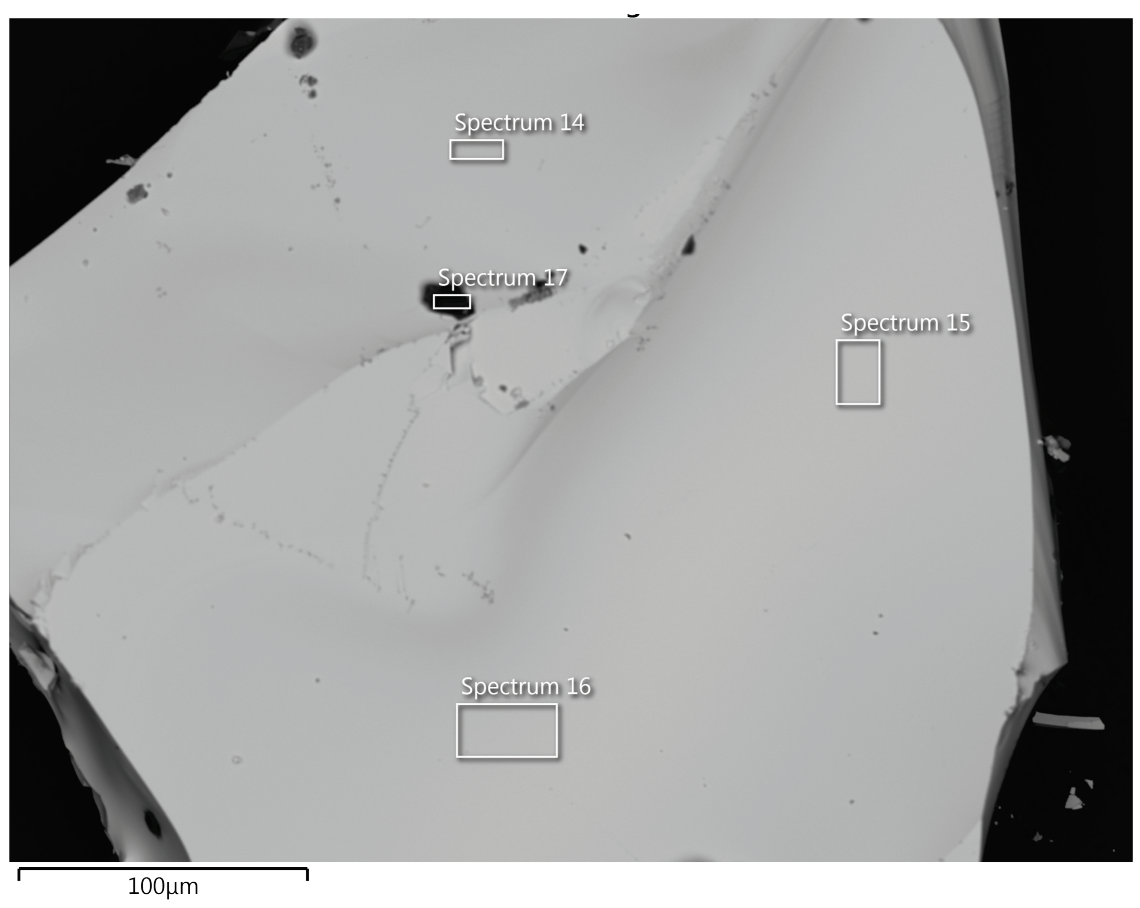

| | At% | standard deviation |
|---|---|---|
| U | 0.057 | 0.001 |
| Ir | 0.207 | 0.003 |
| Al | 0.736 | 0.004 |

**Fig. S1. SEM-EDS results of $UIr_4Al_{15}$.**

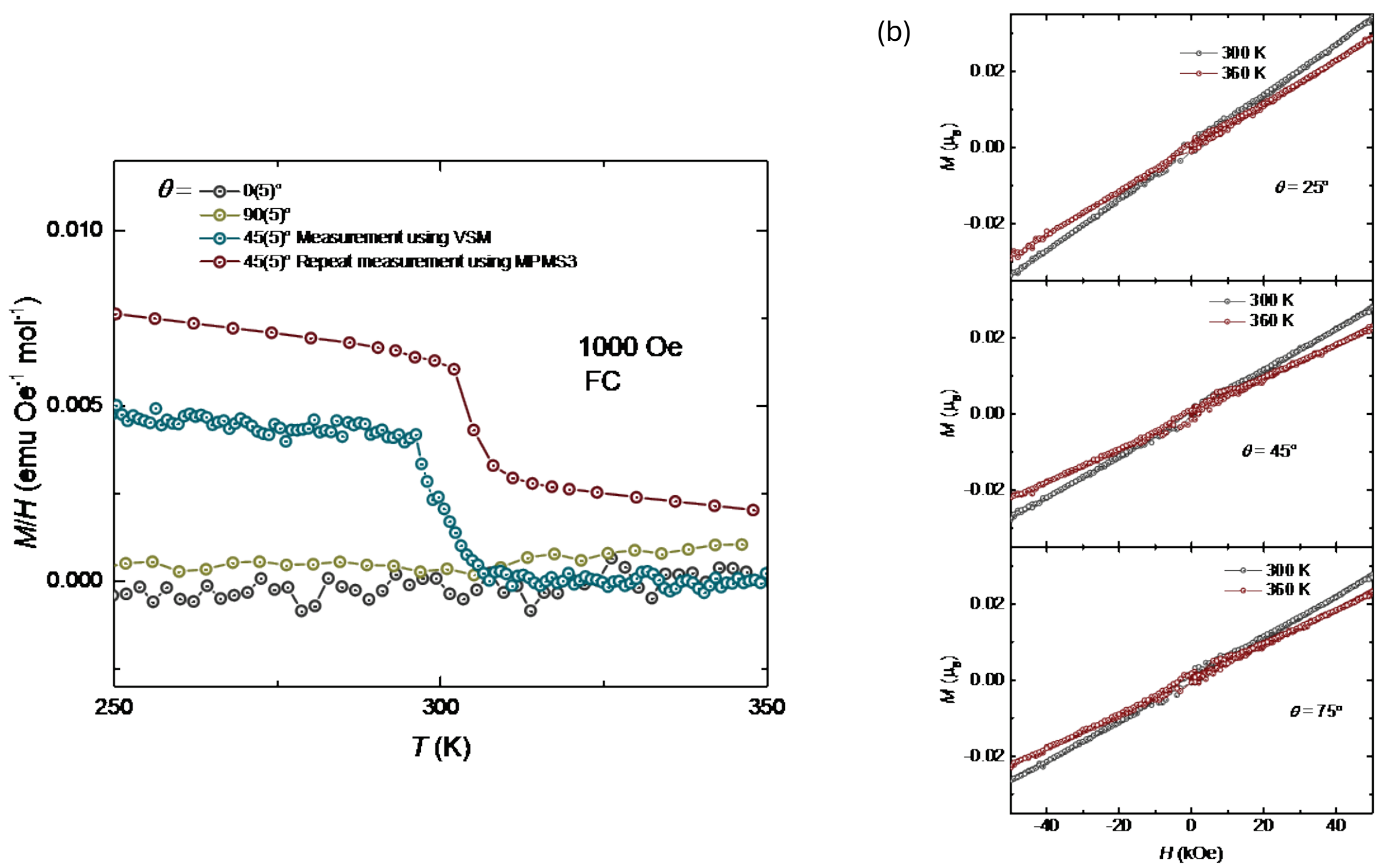

**Fig. S2.** Cross-platform reproduction of the tilted-field magnetic anomaly at 1000 Oe on field-cooled branches. The same crystal was remounted and measured using VSM temperature-sweep mode (2 K $min^{-1}$) and an MPMS3 with no-overshoot temperature control (2 s wait after temperature stability). The nominal tilted orientation is approximately 45° from c; its estimated uncertainty is 5–10°. Reference measurements near the principal directions are also shown. No vertical offsets or background subtraction were applied. Both platforms reproduce a decrease in M/H near room temperature, although their absolute values and apparent anomaly temperatures differ. This comparison demonstrates qualitative reproduction after remounting, not quantitative agreement between platforms.

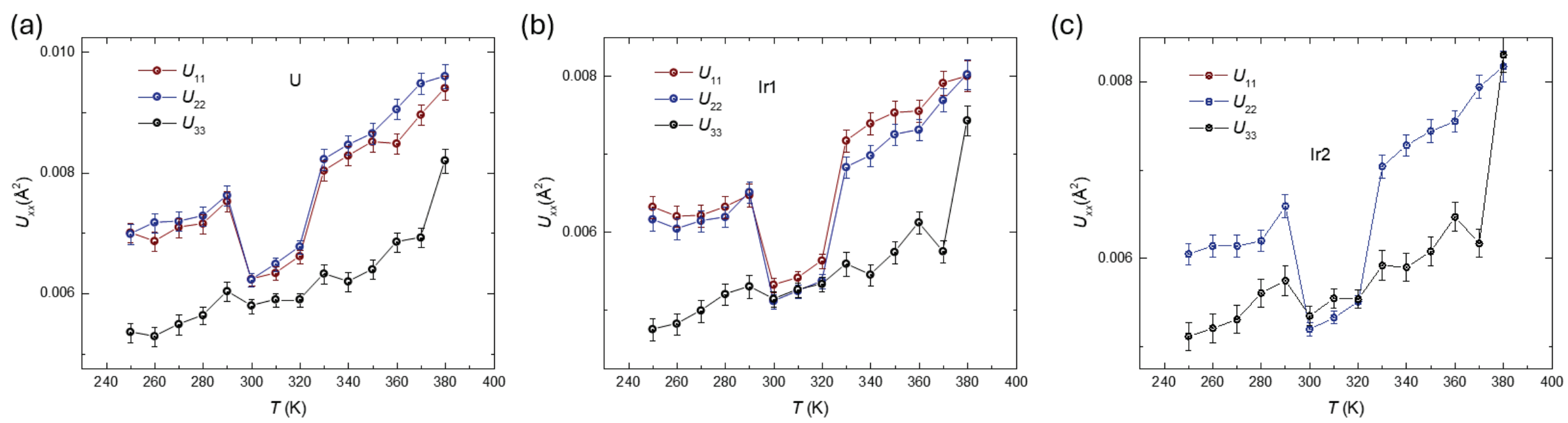

**Fig. S3.** The anisotropic displacement parameters.

## Reference

(1) Canfield, P. C. New Materials Physics. *Reports on Progress in Physics* **2020**, *83* (1), 016501. https://doi.org/10.1088/1361-6633/ab514b.

(2) LSP INDUSTRIAL CERAMICS. Canfield Crucible Sets. **2015**.

(3) Canfield, P. C.; Kong, T.; Kaluarachchi, U. S.; Jo, N. H. Use of Frit-Disc Crucibles for Routine and Exploratory Solution Growth of Single Crystalline Samples. *Philosophical Magazine* **2016**, *96* (1), 84–92. https://doi.org/10.1080/14786435.2015.1122248.

(4) Parkin, S.; Moezzi, B.; Hope, H. XABS 2: An Empirical Absorption Correction Program. *J. Appl. Crystallogr.* **1995**, *28* (1), 53–56. https://doi.org/10.1107/S0021889894009428.

(5) Walker, N.; Stuart, D. An Empirical Method for Correcting Diffractometer Data for Absorption Effects. *Acta Crystallogr. A* **1983**, *39* (1), 158–166. https://doi.org/10.1107/S0108767383000252.

(6) Sheldrick, G. M. SHELXT – Integrated Space-Group and Crystal-Structure Determination. *Acta Crystallogr. A Found. Adv.* **2015**, *71* (1), 3–8. https://doi.org/10.1107/S2053273314026370.

(7) Sheldrick, G. M. Crystal Structure Refinement with SHELXL. *Acta Crystallogr. C Struct. Chem.* **2015**, *71* (1), 3–8. https://doi.org/10.1107/S2053229614024218.

(8) Dolomanov, O. V.; Bourhis, L. J.; Gildea, R. J.; Howard, J. A. K.; Puschmann, H. OLEX2 : A Complete Structure Solution, Refinement and Analysis Program. *J. Appl. Crystallogr.* **2009**, *42* (2), 339–341. https://doi.org/10.1107/S0021889808042726.